\documentclass[conference]{IEEEtran}
\IEEEoverridecommandlockouts
\usepackage{cite}
\usepackage{amsmath,amssymb,amsfonts}
\usepackage{algorithmic}
\usepackage{graphicx}
\usepackage{caption}
\usepackage{subcaption}
\usepackage{textcomp}
\usepackage{xcolor}
\usepackage{layouts}
\usepackage{makecell}
\usepackage{pifont}
\usepackage{soul}
\newcommand{\cmark}{\ding{51}}%
\newcommand{\xmark}{\ding{55}}%
\makeatletter
\newcommand{\linebreakand}{%
  \end{@IEEEauthorhalign}
  \hfill\mbox{}\par
  \mbox{}\hfill\begin{@IEEEauthorhalign}
}
\def\BibTeX{{\rm B\kern-.05em{\sc i\kern-.025em b}\kern-.08em
    T\kern-.1667em\lower.7ex\hbox{E}\kern-.125emX}}
\begin{document}

\title{ Toward Reusable Science with Readable Code and Reproducibility}

\author{\IEEEauthorblockN{Layan Bahaidarah}
\IEEEauthorblockA{\textit{The College of Engineering} \\
\textit{Boston University} \\
Boston, MA, USA \\
layanb@bu.edu}
\and
\IEEEauthorblockN{Ethan Hung}
\IEEEauthorblockA{\textit{The College of Engineering} \\
\textit{Boston University} \\
Boston, MA, USA \\
ehung@bu.edu}
\and
\IEEEauthorblockN{Andreas Francisco De Melo Oliveira}
\IEEEauthorblockA{\textit{The College of Engineering} \\
\textit{Boston University} \\
Boston, MA, USA \\
andoliv@bu.edu}
\linebreakand
\IEEEauthorblockN{Jyotsna Penumaka}
\IEEEauthorblockA{\textit{The College of Engineering} \\
\textit{Boston University} \\
Boston, MA, USA \\
jyotsnap@bu.edu}
\and
\IEEEauthorblockN{Lukas Rosario}
\IEEEauthorblockA{\textit{The College of Engineering} \\
\textit{Boston University} \\
Boston, MA, USA \\
lukasr@bu.edu}
\and
\IEEEauthorblockN{Ana Trisovic}
\IEEEauthorblockA{\textit{Institute for Quantitative Social Science} \\
\textit{Harvard University}\\
Cambridge, MA, USA \\
anatrisovic@g.harvard.edu}
}

\maketitle

\begin{abstract}
An essential part of research and scientific communication is researchers' ability to reproduce the results of others. While there have been increasing standards for authors to make data and code available, many of these files are hard to re-execute in practice, leading to a lack of research reproducibility. This poses a major problem for students and researchers in the same field who cannot leverage the previously published findings for study or further inquiry. To address this, we propose an open-source platform named RE3 that helps improve the reproducibility and readability of research projects involving R code. Our platform incorporates assessing code readability with a machine learning model trained on a code readability survey and an automatic containerization service that executes code files and warns users of reproducibility errors. This process helps ensure the reproducibility and readability of projects and therefore fast-track their verification and reuse.
\end{abstract}


\section{Introduction}

Over the last decade research has become more computational and increased the value of software as a primary scientific asset. A 2014 survey shows that roughly 90\% of UK academics used software in their research~\cite{hettrick2014uk}, and a 2017 survey found that 95\% of US postdoctoral scholars used research software~\cite{nangia2017track}. Researchers do not only use existing software, but they develop new software and write code to carry out their studies. Software plays a critical role in modern research and is increasingly recognized and published as a valuable scholarly contribution~\cite{katz2018publish}. 

While a scientific finding can be clearly described in publications to provide a gist of its research methodology, only original code provides enough detail to allow for research replication~\cite{smith2016software}. By examining the research software used in a published study, one can gain a deeper understanding of the study and its results. Sharing code enables a higher level of research transparency as anyone can review, examine and scrutinize published results in detail~\cite{national2019reproducibility, barnes2010publish, ince2012case}. 
Furthermore, leveraging previously published findings and reusing shared research data and code is critical for further scientific inquiries and reducing the cost of repeating work. 

For the above reasons, it is clearly highly beneficial to the scientific community as a whole to make the published research code transparent and reusable. However, in practice, many researchers develop code with a primary goal of obtaining scientific insights and its reuse aspect is often neglected. Additionally, code development is usually carried out by early-career researchers who may lack formal programming education or training~\cite{hannay2009scientists}. Due to this lack of training and placing priority on other aspects of research, published code can often be poorly formatted and documented, and thus unintelligible for students and researchers who aspire to reuse it. Even re-executing published code files can be troublesome~\cite{trisovic2021large, pimentel2019large}, which hinders the prospect of research reproducibility, defined as the ability to obtain consistent results using the same input data, code, computational steps, and methods~\cite{national2019reproducibility}. This is problematic, given that reproducibility is an essential part of the scientific process.

To understand how to improve the current situation, it is useful to note that software development in academia is not as well-supported as it is in industry. For the most part, software developers in the industry receive training, and their code is reusable and extensible because it commonly follows a required style and good practices. An extensive literature has been produced for the developers on clean code and standards~\cite{hyde2020write, thomas2019pragmatic, martin_clean_2009, van2008software}. For instance, code readability is recognized as an important factor in code quality and transparency. In contrast, current software standards in academia have been primarily focused on its dissemination and open-source values~\cite{smith2016software, lamprecht2020towards, jimenez2017four}. In this paper we will instead focus on its usability from a reuser's perspective by facilitating its readability and re-execution.

One strategy for how to improve the code reusability is to increase peer or community review. However, a significant amount of published code is not reviewed~\cite{hafer2009assessing}, often due to niche applications or lack of resources. Therefore, academia should strive toward solutions that can automatically evaluate the code before it is published.

This paper presents an approach for improving code reusability using a machine learning model to evaluate its readability and an automatized workflow to test its reproducibility. We conduct a survey of students and researchers to learn what code features are considered desirable and use the results as input to the model. We use virtualization technology for the reproducibility service. Further, the paper presents an implementation of the readability and reproducibility services as an online platform RE3.~\footnote{The platform is available online at re3.ai.} We expect that the services and the results presented in this paper will be of interest to research repositories, researchers who develop code in R, and the communities advancing scholarly communications, research reproducibility, and reuse. 

Our main contributions can be summarized as follows:

\begin{itemize}
    \item We identify code readability as an important factor in reuse, conduct a survey to collect data on code readability, and create a machine learning model that predicts code readability based on its features. 
    \item We conduct an analysis to identify code features that correlate to code readability. We use our machine learning model to rate real-world research code over the years to examine how its readability has changed.
    \item We propose an automatized approach for testing research reproducibility, where the code is tested in a clean environment upon its submission for publication. If the code does not execute correctly, the researcher receives feedback to help fix the error.
    \item We develop an interactive lightweight platform that incorporates code readability and reproducibility services. 
    \item Our data, model, and code are released open access and can be freely reused.
\end{itemize}

The rest of the paper is organized as follows. Section~\ref{sec:read} provides a detailed description of the code readability survey, data analysis, and our machine learning model. Section~\ref{sec:repro} explains our implementation for reproducibility testing. We present the implementation of the services and the online platform RE3 in Section~\ref{sec:app}. Section~\ref{sec:diss} describes how we validated our platform using real-world research data and code from the Harvard Dataverse repository. We also discuss the primary use-cases and limitations of our services. Section~\ref{sec:related} describes related work. Finally, Section~\ref{sec:conc} concludes the paper considering future work.

\section{Code readability}\label{sec:read}

\begin{figure}[htbp]
\centerline{\includegraphics{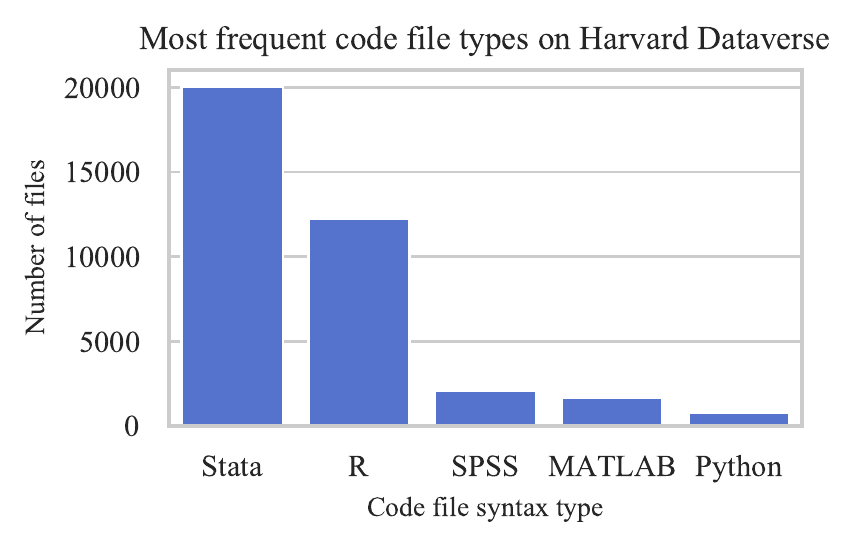}}
\caption{Bar plot of the most frequent code file types stored in Harvard Dataverse. Adapted from Ref.~\cite{DVN/UZLXSZ_2021}.}
\label{fig:langs}
\end{figure}

We use the R programming language for the implementation of our code readability model. R is frequently used for statistical computing and is among the most popular languages in academia and data science~\cite{vilhuber2020report, kaggle}. The Harvard Dataverse research repository stores over 2,500 replication datasets with more than 10,000 unique R code files published with scientific data (see Fig.~\ref{fig:langs}). While Stata is the most popular type of code file on Harvard Dataverse, it is proprietary and its implementation on the cloud would require a licence. R is an ideal candidate for our implementation because it is popular, free and open source.

\subsection{Code Rating Survey}

In order to develop a machine learning model for code readability, we organized a survey to collect ratings of 100 unique R code snippets. The survey would help us determine what code features are considered readable and unreadable. 

As students and early-career researchers are some of the primary reusers of scientific code, the survey was mainly distributed to the students at the Electrical and Computer Engineering department of Boston University. To encourage participation, two Amazon gift cards were raffled to all participants who completed the survey within three weeks of its release. Participants were also encouraged to complete the survey irrespective of their background experience. 

We created a collection of 100 code snippets according to the following procedure. First, all snippets contain 12 to 18 lines of code so that they can be promptly visually evaluated in the survey. Second, the code is selected to be diverse and reflect a variety of features. For example, since average line length is a feature, then the survey snippets should contain high, medium, and low average line length. Third, all selected snippets have a high degree of cohesiveness considering the functionality of the code~\cite{buse2009learning}. Finally, the collection has a balanced mixture of readable and non-readable code. We use 30 canonically "good" snippets from officially maintained R libraries at Tidyverse\footnote{https://www.tidyverse.org} and 70 diverse snippets from Harvard Dataverse\footnote{https://dataverse.harvard.edu} published by researchers.

At the beginning of the survey, all users are asked to complete a series of onboarding questions. These questions were asked to gauge the user's programming experience, years in academia, and familiarity with the R programming language. The users are then redirected to the code rating page, where they are shown the same sequence of 100 syntax-highlighted R code snippets. For each snippet, the user is presented with buttons ranging from 1 (unreadable) to 10 (highly readable), allowing them to rate the readability of each snippet based on their intuition of code readability. Users also have access to a ``previous'' button to change the rating of a previously rated snippet and a ``skip'' button in case they are unsure how to rate it. Lastly, a ``help'' button is included to acknowledge that readability is based on their intuitive notion rather than some predefined concept. Users are not required to complete the entirety of the survey, and the ratings they contributed up until that point are still saved. 

We received a total of 3694 snippet ratings from 68 survey participants, out of which 32 completed the survey in full. Most of the participants (67) had a background in STEM, and 1 had a background in the social sciences. Additionally, 36 people had 0-3 years experience with coding, 20 had 3-5 years, and 12 had more than 5 years of experience. Finally, 56 people had no experience with R, and the remaining 12 did. We found that the average absolute score difference between people with R experience compared to those without was only 0.8. While this is not insignificant, we consider it small enough to suggest that people without R experience, on average, can rate the code just as well compared to participants who do have R experience.  In fact, the R snippets selected as "canonically good" had a greater average rating from users with no R experience than those with it.

\subsection{Machine Learning Readability Model}

The input to our machine learning model consisted of static structural features of R code snippets, including indentation, comments, and line length. These structural features can play a major factor in one's judgment of code readability~\cite{buse2009learning}. Since the snippets were between 12-18 lines of code and unrepresentative in size of real-world code uploaded to research repositories, all features were standardized per number of code lines. Specifically, features were calculated as an average per line or the maximum value for all lines. By doing this, we ensure that our machine learning model generalizes to longer pieces of code rather than just our selected snippets of code. Our list of features is modeled after Ref.~\cite{buse2009learning} but modified to fit the R programming language use case (Tab.~\ref{tab:features}). R keywords include \texttt{if}, \texttt{for}, \texttt{next}, but also \texttt{TRUE}, \texttt{Inf} and \texttt{NA}. Loops are a subset of keywords (\texttt{for}, \texttt{while} and \texttt{repeat}), while branches overlap with a difference (containing \texttt{if}, \texttt{else} but also "\texttt{else if}"). Variables (identifiers) are matched as strings outside brackets (which are "parameters" or "arguments"), quotation marks, and comments that are not keywords. The feature character identifies the most common single character per R file. Comments include only single-line comments, which in R start with \texttt{\#}. We use regex in Python to automatically identify the code features. 

For the dependent variable (target of prediction) of our machine learning model, we take the mean rating for each of our 100 snippets from the readability survey results. Further, we also label snippets as "readable" and "not readable" represented by a 1 or 0 value, respectively.~\footnote{The labels were assigned for comparison purposes with respect to Ref.~\cite{buse2009learning}.} To obtain this split, we denote code scoring a 5 or below as "non-readable" and scores above 5 as "readable".

\begin{table}[htbp]
\caption{The list of considered code features, each of which represents either an average value per line of code or a maximum value for all lines.}
\begin{center}
\begin{tabular}{ |c||c|c|}
\hline
\textbf{Feature Type} & \textbf{Avg. per line} & \textbf{Max. per all lines}\\
\hline
Arithmetic operators & \cmark & \xmark \\ 
Assignments & \cmark & \xmark \\ 
Blank lines & \cmark & \xmark \\ 
Branches & \cmark & \xmark \\ 
Character & \xmark & \cmark \\ 
Commas & \cmark & \xmark \\ 
Comments & \cmark & \xmark \\ 
Comparison operators & \cmark & \xmark \\ 
Indentation & \cmark & \cmark \\ 
Keywords & \cmark & \cmark \\ 
Line length & \cmark & \cmark \\ 
Loops & \cmark & \xmark \\ 
Numbers & \cmark & \cmark \\ 
Parentheses & \cmark & \xmark \\ 
Periods & \cmark & \xmark \\ 
Spaces & \cmark & \xmark \\
Variables & \cmark & \cmark \\
\hline
\end{tabular}
\label{tab:features}
\end{center}
\end{table}

\begin{figure}[htbp]
\centerline{\includegraphics{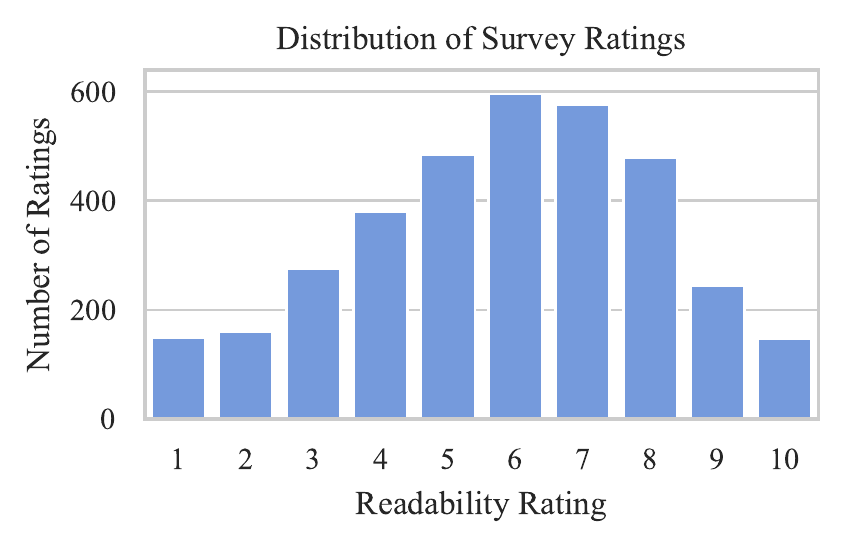}}
\caption{Distribution of readability ratings from all users who participated in the survey (total of 3694 ratings).}
\label{fig:rdist}
\end{figure}

The distribution of the survey ratings is shown in Fig.~\ref{fig:rdist}. The rating average is 5.8, while the median is 6. The original ratings were kept with the exception of cases where ratings significantly diverged from other users. These outliers, defined as a rating 1.5 times the interquartile range above the upper quartile and below the lower quartile, were identified and removed for each R code snippet. A total of 59 user ratings (1.6\% of all ratings) were removed from the dataset. Ultimately, our final model excluded the outlier ratings.

\begin{figure}[htbp]
\centerline{\includegraphics[width=\linewidth]{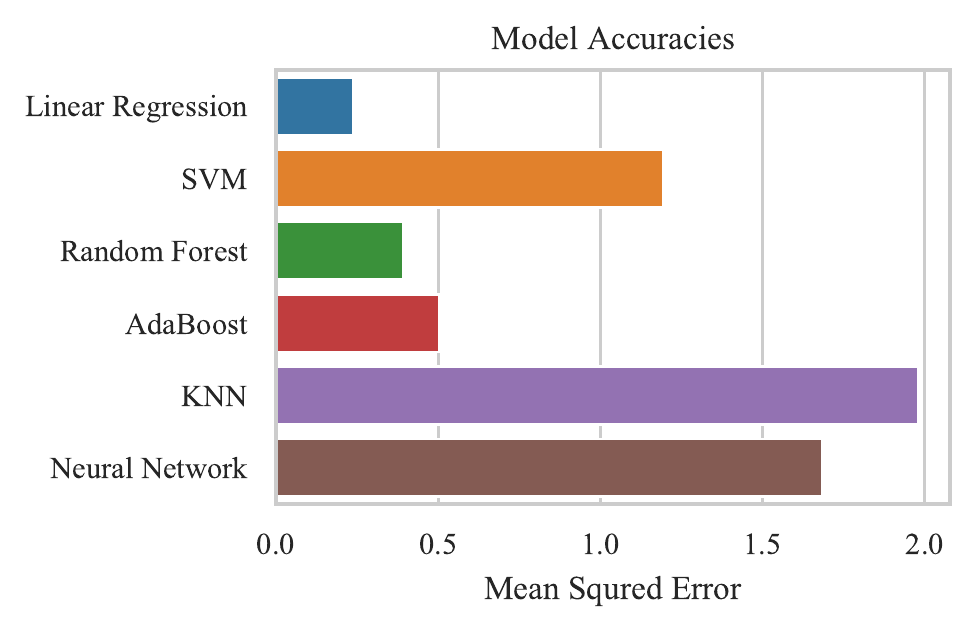}}
\caption{Performance of ML models.}
\label{fig:mse}
\end{figure}

Following the extraction of the code features, collection of the survey ratings, and data cleaning, we tested a number of machine learning models. We focus on regression but also consider the classification task to compare with prior results in the field. Building upon recent ML advances, we tested the following models: SVM, Random Forests, AdaBoost,  Neural Networks, Naive Bayes and KNN. The models were trained on 80\% of randomly selected code snippets, and each trained model was cross-validated ten times on the training dataset to ensure it was not overfitting. Moreover, models that had hyperparameters were trained on a variety of different settings in order to extract an approximate best hyperparameter set. The mean squared error incurred by each model in the regression task is shown in Fig~\ref{fig:mse}, with the linear regression model achieving the best testing result. This model was then selected to be incorporated into our platform in Section~\ref{sec:app}. 

After testing a variety of models on the classification task by working with binary ratings (1 if a readability score is higher than 5 and 0 otherwise), we achieve an accuracy of 85\%, 85\% and 90\% with Logistic Regression, AdaBoost and Random Forest, respectively. This is comparable to Buse and Weimer~\cite{buse2009learning}, who report an accuracy of 80\%. 

For comparison and validation, we ran our model on data collected and published by Buse and Weimer~\cite{buse2009learning} to evaluate how well R readability corresponds to Java readability. We parsed all 100 Java snippets with the same feature parser used in our original R model. While the R and Java syntax have some similarities like the use of braces and (partially) variable assignment, they differ in variable definition of built-in data types, function definition, and some of the language keyword (notably \texttt{TRUE} in R is \texttt{true} in Java). Therefore, we are able to obtain only an approximate set of features for the Java language. We then run our model on the generated features and compare their scores to those collected from Buse and Weimer's survey. Doing so, we reached an accuracy of 59\%, signifying that the model has some predictive ability on Java code despite not using a Java feature parser. The output could be improved with the use of a feature generation parser for Java.

The inter-rater agreement statistics for human versus human annotators suggest that there was a moderate agreement for any given snippet on how well it should be rated (Tab.~\ref{tab:stests}). Additionally, the p-values for the average human versus human annotator agreement are all greater than 0.05, indicating that the raters are statistically independent (as they should be). The inter-rater agreement statistics for the prediction made by the model versus the human is slightly higher than the human versus human, which is expected. Also, the p values for the model vs. human statistics are all below or equal to 0.05, indicating the two are not independent, which is also expected since the model is based on the human survey data.

\begin{table}[htbp]
\caption{Statistical tests}
\begin{center}
\begin{tabular}{|c|c|c|}
\hline
\textbf{Statistical Test} & \textbf{\thead{Avg. Human-Human\\ Annotator Agreement}} & 
\textbf{\thead{Avg. Model-Human\\ Annotator Agreement}} \\
\hline
Spearman's $\rho$ &  0.196 (p=0.185) & 0.333 (p=0.05)\\
Pearson's $r$  & 0.226 (p=0.151) &  0.39 (p=0.042)\\
Kendall's $\tau$ & 0.156 (p=0.181) & 0.283 (p=0.054) \\
\hline
\end{tabular}
\label{tab:stests}
\end{center}
\end{table}

\subsection{Feature Predictive Power}

\begin{figure}[htbp]
\centerline{\includegraphics{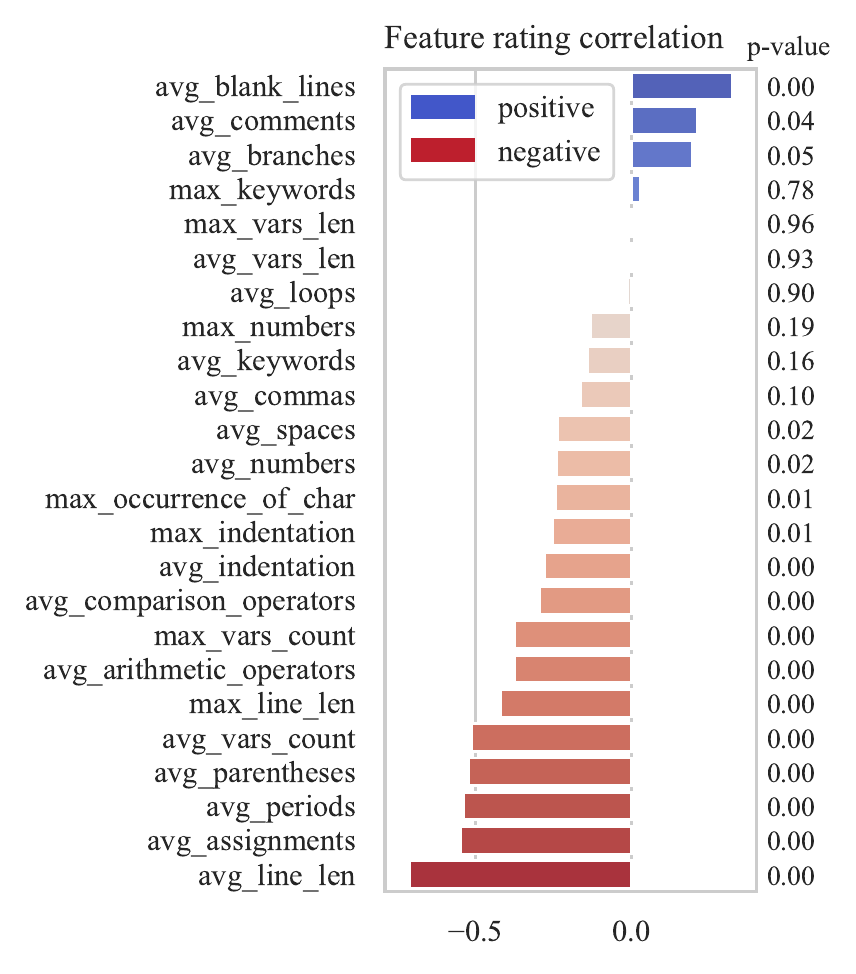}}
\caption{Feature predictive power.}
\label{fig:feat}
\end{figure}

Feature importance is a model inspection technique that allows us to evaluate how much the model depends on each input feature. In other words, features that had the most impact on our model are perceived as most important for code readability. Features with their correlation with the model and p-values are shown in Fig.~\ref{fig:feat}. We found average line length to play the most significant factor in our model, meaning that code containing long lines negatively affects readability. Features such as numbers, keywords, and commas have little impact on the model. Comments in the code have a relatively low positive correlation coefficient, which may be counterintuitive as it is often taught as a best practice. It suggests that commenting can be a misunderstood practice and that, while one should comment their code, adding comments does not necessarily improve unintelligible code. Existing software engineering recommendations propose using a limited number of comments and emphasize that comments in the code should be avoided with intuitive variable naming~\cite{martin_clean_2009}, which corroborates our finding.

\subsection{Readability ratings of real-world research code}

\begin{figure}[htbp]
\centerline{\includegraphics{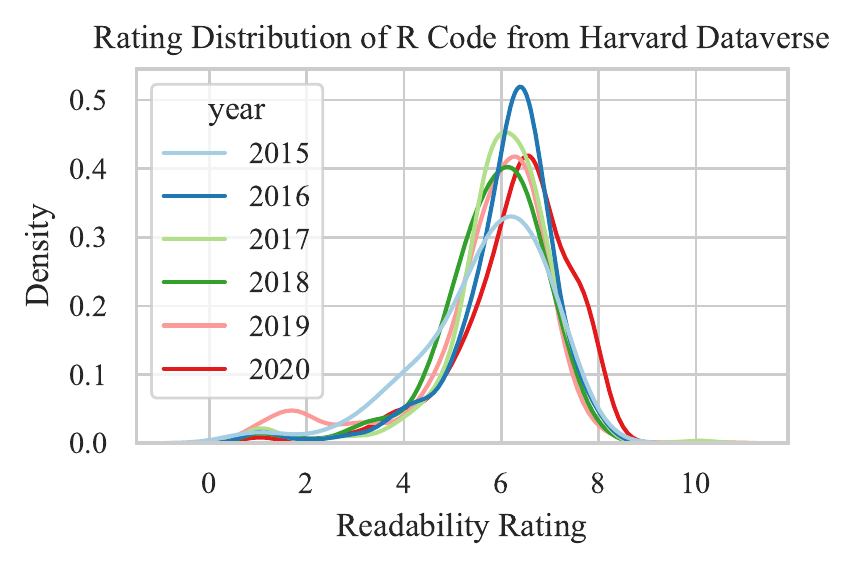}}
\caption{Distribution of readability scores for over 1700 unique R code files shared on the Harvard Dataverse repository.}
\label{fig:dvcode}
\end{figure}

We use our model to rate 1740 real-world R code files from the research data repository Harvard Dataverse. The files were published as replication packages and the distributions of readability scores per year are shown in Fig.~\ref{fig:dvcode}. We can see that most files obtained a readability score between 5.0 to 7.0 and that readability scores ranged between 1 to 10. We could expect that with time, published code is held to a higher standard and that because of it and new specific training, the code would appear 'more readable' with years. However, we do not observe a significant improvement in the code readability over the last five years, with two exceptions. In 2015 there is a higher density of scores between grades 2 and 4, suggesting presence of code with low readability, while 2020 has a a higher distribution of scores above 7, not present in the prior years, which is an indicator of code with higher readability.

\section{Service Application and the RE3 Platform}\label{sec:app}

This section describes the implementation of the RE3 web platform that incorporates code readability and reproducibility services. The platform currently offers three choices: participation in the readability survey (as described in Section~\ref{sec:read}), a 'stand-alone' code readability test, and research sharing. The code readability test has an independent page that rates code from a text box in the browser and provides recommendations for improvement. The research sharing choice enables users to upload their data and code, and it incorporates both reproducibility tests and readability ratings. 

\subsection{Technical implementation}

The RE3 platform is developed on the Google Firebase.\footnote{https://firebase.google.com} It is an application development software that provides tools for tracking analytics, web hosting, debugging, and product experiments. For file storage and database services, we use Cloud Storage for Firebase and Cloud Firestore, respectively. Firestore, which is a NoSQL database native to the Google ecosystem, captures the metadata of all files and projects on the platform. Last, we use Flask, a Python microframework for web serving and testing. Python is used as the primary programming language for all development on the platform.

\subsection{Research sharing workflow}

\begin{figure*}[htbp]
\centering
\begin{subfigure}{.5\textwidth}
  \centering
  \includegraphics[width=.8\linewidth]{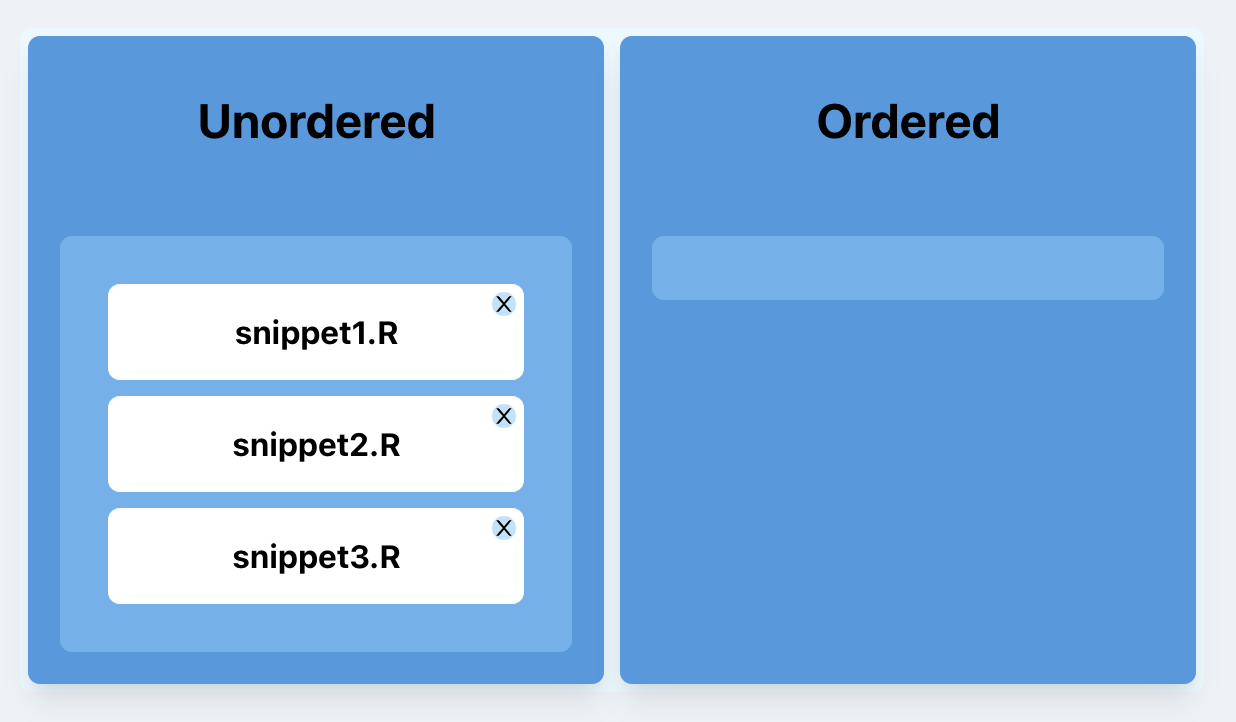}
  \caption{The code files have been uploaded.}
  \label{fig:sub1}
\end{subfigure}%
\begin{subfigure}{.5\textwidth}
  \centering
  \includegraphics[width=.8\linewidth]{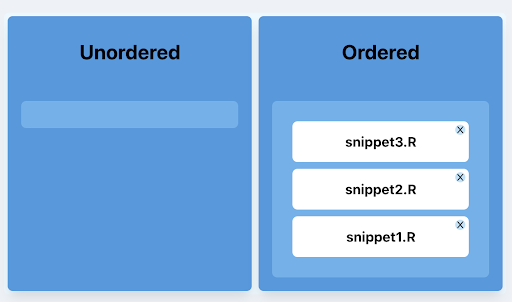}
  \caption{After the user specifies the order.}
  \label{fig:sub2}
\end{subfigure}
\caption{An interface for specifying the execution order for uploaded code files.}
\label{fig:order}
\end{figure*}

Generally speaking, data and code sharing is one of the last research processes, and the researchers often do not prioritize spending much effort on it. Therefore, it is important to make this process easy and fast while also obtaining the necessary information about the data and code. Considering that likeability and ease of use are critical factors in adopting a service~\cite{teo2003empirical}, a functional yet simple UI has been implemented for the RE3 web platform.

Upon navigating to the research sharing page, users are prompted with a form collecting basic information about the project, including:

\begin{itemize}
    \item Author name
    \item Title of the research project
    \item Required R version 
    \item Code license
    \item Data license
    \item Keywords
\end{itemize}

The captured information is used as the project metadata or a description of uploaded resources used to enable their findability, reusability, and assessment. The FAIR principles, which are high-level interdisciplinary practices, advocate for the use of standardized machine-readable metadata to make data findable, accessible, interoperable, and reusable~\cite{wilkinson2016fair, jacobsen2020fair}. The first two fields (author and title) together with the publication date are citation-level metadata. As the platform is currently geared toward the programming language R, we ask the researcher about the used R software version. Our form features both data and code license fields as different types of licenses apply to these resources~\cite{jimenez2017four}. The absence of one of the two licenses in a replication package with data and code may create legal uncertainty. Though the licenses should be open to allow reuse, no license within the published material means that all copyright laws apply and that the author retains all rights to data or code. To create a platform that facilitates reusable research, licenses for both types of resources needed to be incorporated. Last, the keywords facilitate project findability.

After completing the form, the users upload their data and code. In some cases, the research code files are meant to execute independently of each other, but in others, they are supposed to be executed in a specific sequence (workflow). The platform features a drag-and-drop component, as shown in Fig.~\ref{fig:order}, to help the users specify the execution order. All uploaded code files will appear in the unordered column, which the users can 'drag' to the ordered column. At the upload, all code files will be rated on readability and tested for reproducibility. The readability scores are automatically generated with our machine learning model for all R code files. The project reproducibility is tested as described in Section~\ref{sec:repro}. On successful code re-execution, the generated files are displayed as ``Project Artifacts''. Code reviewers and reusers can then compare the generated plots with the expected results. 

Gamified components are incorporated into the service to improve user engagement~\cite{huotari2017definition}. The user receives color-coded readability scores of their code, and if the score is low, it should incentivize the user to increase their score by improving their code. Further, a suggestion box will be displayed to help the user improve their code's readability to obtain a better score. The suggestion box supplies the user with some of our key findings on the features that correlate with readability scores. It will also suggest other good practices, like adding comments, if no comments were detected in the code. Both readability testing and containerizing a project might require multiple iterations, depending on the initial code quality. Finally, the updated researcher's code, together with the container, are published on the platform. 

\section{Reproducibility service}\label{sec:repro}

One of the major challenges in computational reproducibility and reuse is capturing code dependencies necessary for its execution. Code dependencies can be software meant to build or run the code, external libraries, and system requirements needed to run the code correctly. These dependencies are currently not captured for much of the published research code~\cite{trisovic2021large, pimentel2019large}, which is one of the major causes of non-reproducibility reported across fields~\cite{national2019reproducibility}.   

One of the best reproducibility practices is to test research code in a clean environment before publishing it or sharing it online. Doing that helps the researcher (or code depositor) identify common errors such as fixed file paths, missing dependencies, and missing files. However, in practice, that is rarely happening, and the code is deposited untested. Nevertheless, research code could become more reusable if code and data repositories themselves incorporate and support testing the code at upload.

\begin{figure}[htbp]
\centerline{
\includegraphics[width=\linewidth]{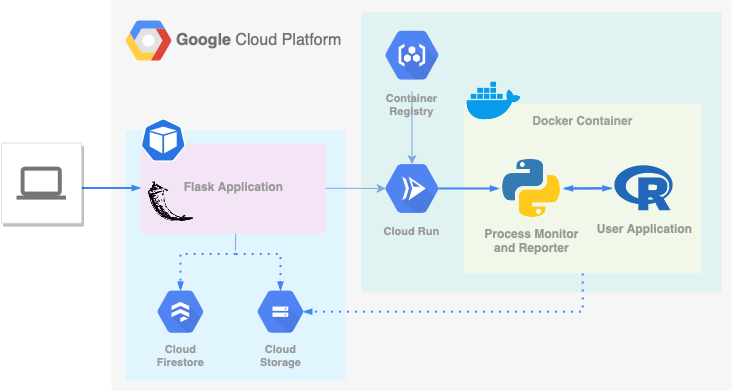}}
\caption{Schema (simplified) of the reproducibility service. For a detailed diagram of the back-end, see Appendix.}
\label{fig:repro}
\end{figure}

We propose a lightweight service that executes the code at the upload to support reproducibility (Fig.~\ref{fig:repro}). First, the user uploads project files and associated information to RE3.ai, which asks them to indicate the used software (R version). The web application sends the code, data and dependency information to Cloud Build, a Google-managed service for building Docker images. Docker has played a significant role in the effort for reproducibility~\cite{jimenez2015role, cito2016using}. For the testing step, we use the Docker virtualization technology that bundles software, libraries, and configuration files together and runs them as containers. During the build process, Docker creates a virtual environment using Conda, installs the R version specified by the user, retrieves the researcher's code files, executes them in the order specified by the user during the upload step, and records and outputs the execution result, which can be success or error. All configurations for the Docker containers, the code and data files are saved to ensure a reusable, standardized environment that can guarantee the code will execute as intended in the future. If we encounter any errors during the process of containerization or from running the code, the researcher will receive feedback and will be able to update their code to resolve the error. As a result, we ensure that uploaded code is executable and reproducible by design.

\section{Evaluation and Discussion}\label{sec:diss}

For the purposes of investigation and evaluation, we retrieved 207 real-world replication packages from the Harvard Dataverse repository. All packages amounted to less than 10~MB, contained code files (only in the R language), and were published from 2016 to 2020. A quick text-parsing analysis revealed that the used R version was documented in deposited files in the format "R x.y.z" in 44 out of 207 (21\%) replication packages. Considering that the R version is essential for successfully re-executing old code files, this outcome is quite low. In the RE3 platform, we directly ask the depositor to specify the version as part of the submission system.

In order to evaluate the reproducibility service, we attempt to containerize three arbitrarily chosen replication packages from Harvard Dataverse.
The first package~\cite{DVN6QE7NN_2019} did not document the used R version, so we used R 3.6, one of the most stable. The code from the package re-executed on our platform without any errors. The second~\cite{DVN4C1KJT_2020} replication package documented used R version (3.5.1) directly in the code files. However, its initial re-execution threw an error, pointing that a library was missing.  We modified the code to install and load the missing library, which led to its successful re-execution. The third project~\cite{DVNBPU1XG_2018} documented the used R version (3.4.1) inside a README file. Its code files re-executed without any errors. Therefore, we were able to build and run all three projects in Docker containers from our platform. Our platform provided all specified R software versions and facilitated making changes in the source code to enable re-execution. We can conclude that it can successfully run and containerize real-world projects.

All in all, we identified a lack of resources for research code review and reproducibility testing as a gap in the scholarly ecosystem and developed automatized services that could fill this gap. Transparency of research code is necessary for its reuse, and it can be aided with higher readability. Supporting re-executable code and reproducible results within the sharing infrastructure could help relieve the reported reproducibility crisis~\cite{national2019reproducibility}. While our reproducibility service currently supports analyses created for a single computer, the readability service can be used on any R code created for all infrastructures, from personal to supercomputers. Though our platform is currently a prototype implementation, it shows the operational capabilities of our services for many use-cases addressed in data and code repositories. 

The RE3 readability and reproducibility services could be incorporated in many ways to automatically improve shared code in research and code-sharing infrastructures. First, the reproducibility test and the readability score can be a barrier to entry before uploading any code to research repositories, which would ensure that all deposited code is tested. In the case that a barrier of entry is too strict, another possibility is to warn the user if the code they are uploading to the platform has obtained a low readability score or is not executable. While the researcher may choose to ignore this warning, it could also encourage them to spend additional time cleaning up, commenting, and formating code before uploading. Last, this feature can also be used as a searchable attribute for research repositories to filter out research code with low readability scores. This could prevent students and researchers from wasting time trying to understand unformatted, uncommented, and unclear code.

In addition to the research-sharing infrastructures, our services could be independently used as a part of the submission system for academic journals. Journals now increasingly encourage data and code sharing as part of the publication, and some carry out code review and result verification~\cite{vilhuber2020report}. Our readability and reproducibility services could accelerate the process of review for the journals by making the code more readable and executable by design.

\begin{figure*}[htbp]
\centering
\centerline{\includegraphics[width=.8\textwidth]{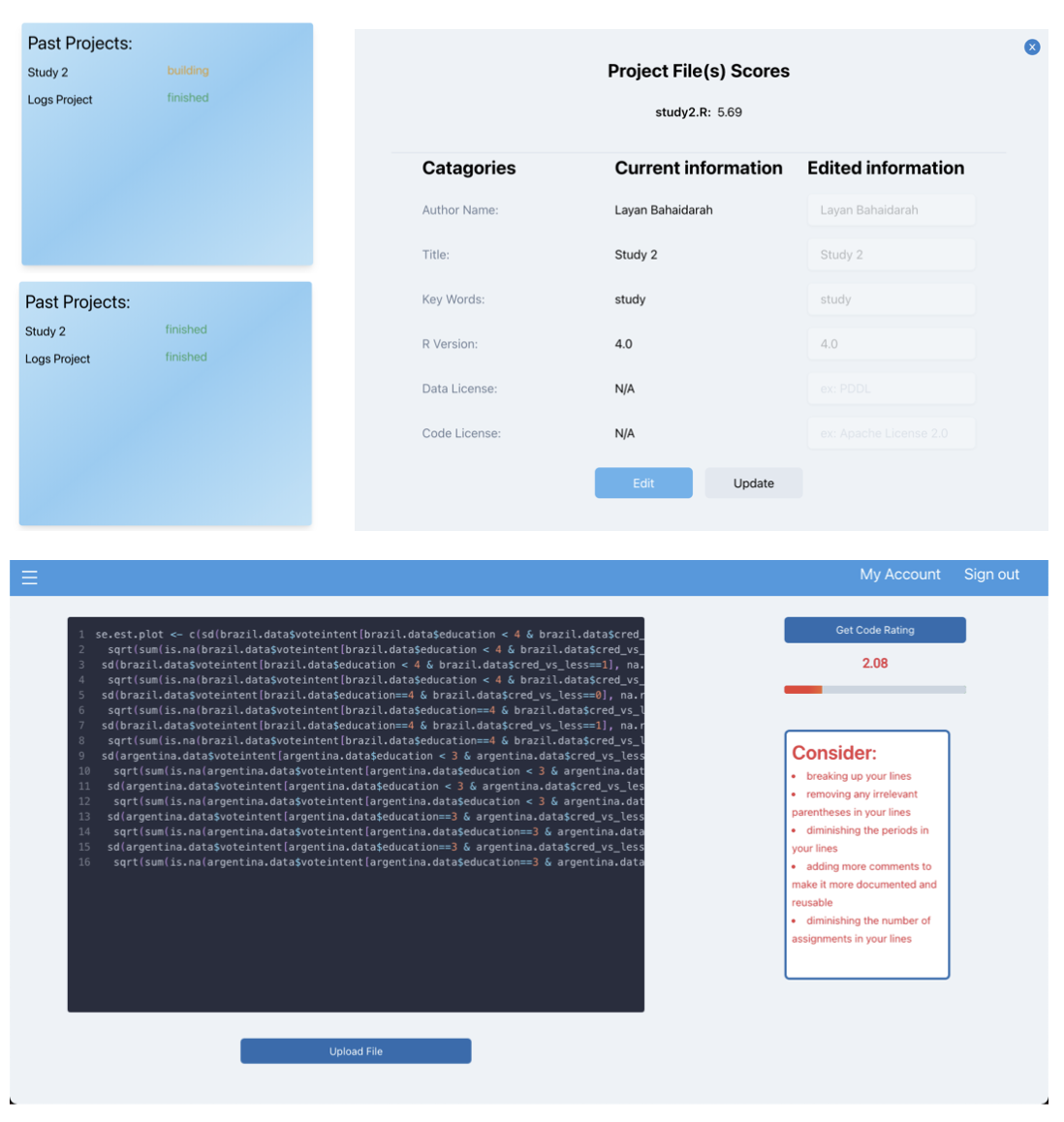}}
\caption{User interface and views of the RE3 platform.}
\label{fig:feedback}
\end{figure*}

\section{Related Work}\label{sec:related}

In this section, we review work related to both readability metrics for code and reproducibility web services (reproducibility platforms).

\subsection{Code Readability}

The readability part of our effort was based on the pioneering paper by Buse and Weimer~\cite{buse2009learning}. They introduced an automated code readability metric using data collected from human annotators. While their study focused on Java code, we aimed for R due to its popularity in academia. Also, the code snippets used in the original study were canonically good (system code), while we used a diverse mix of research and professional code. The accuracy of our models, which used recent advances in machine learning, exceeded the report one. However, there have been significant studies building on the work of Buse and Weimer. Prominent examples include Posnet~\cite{posnett2011simpler}, who introduced a model based on entropy, code size, and Halstead's metrics, while Dorn~\cite{dorn2012general} introduced novel features related to the visual and linguistic presentation of code. Dorn performed a large-scale code readability survey on Java, Python, and CUDA code that included 5000 human annotators. Both of these studies, including others, report an increase in the model accuracy. However, neither of the papers (to our knowledge) reported on the R readability nor applied the models in the research infrastructure as we propose.

\subsection{Reproducibility platforms}

Reproducibility platforms like Code Ocean~\cite{clyburne2019computational, cheifet2021promoting} and Whole Tale~\cite{brinckman2019computing} have been developed to meet the demand for creating reproducible research. They represent a huge undertaking and essentially provide an integrated development environment (IDE) directly from the browser. The users can upload the code, test, and make changes directly in the browser before publishing it online. Both tools encourage the researchers to create a new file (in bash command language) to specify the execution order for their code. On our platform, we use drag-and-drop ordering to encourage the user to specify the sequence. Our approach is more interactive and intuitive, especially for users who are unfamiliar with bash scripts. Further, while the integration between reproducibility platforms and data repositories is feasible~\cite{trisovic2020advancing}, our automatic and lightweight approach for code readability and reproducibility can be easily adopted by all research infrastructures that include an upload web form for their users.

\section{Conclusion}\label{sec:conc}

Without adequate support in the scholarly ecosystem for publishing working research code, we can expect that the number of computationally non-reproducible studies will grow with the growing use of software in the community. Similarly, without code review or readability assessment, new code deposits in these research repositories may continue to be unformatted and hard to follow. To aid in the issue, this paper introduced services for verifying the reproducibility and readability of research projects before they are published online. Through our web platform and a reproducibility pipeline, we showed that it is possible to ensure R code re-execution while asking the user minimal input information such as R software version and a code license. The platform also produces a code readability score and recommendations for improving its readability and documentation. The readability machine learning model was trained on over 3000 human code ratings obtained in a survey. The code recommendations are based on features that highly correlate with the code readability rating.

There are several natural avenues for future work. First, the code readability survey could be a continuous component of the platform allowing for an adaptive readability model. In other words, the readability machine learning model could be continually improved using the new entries from the survey on the platform. As a result, model accuracy would be improved while also capturing new code formatting and functionality trends. Second, given that the platform already incorporates readability ratings, a future version could include a leaderboard based on the reusability scores of users' projects. That way, the users with the tidiest code would be awarded increased visibility at the leaderboard's top spots. Third, RE3 is custom-made for projects involving the R programming language. In the future, this functionality can be extended to support other popular languages, particularly the open-source ones like Python, Julia, or Java. Last, the code re-execution is a necessary prerequisite but not sufficient to guarantee reproducibility. Therefore, our reproducibility service could potentially verify the code files past their re-execution, including running code tests and comparing code outputs such as logs, tables, or plots to the expected results.


\appendix

\begin{figure*}[htbp]
\centering
\centerline{\includegraphics[width=\textwidth]{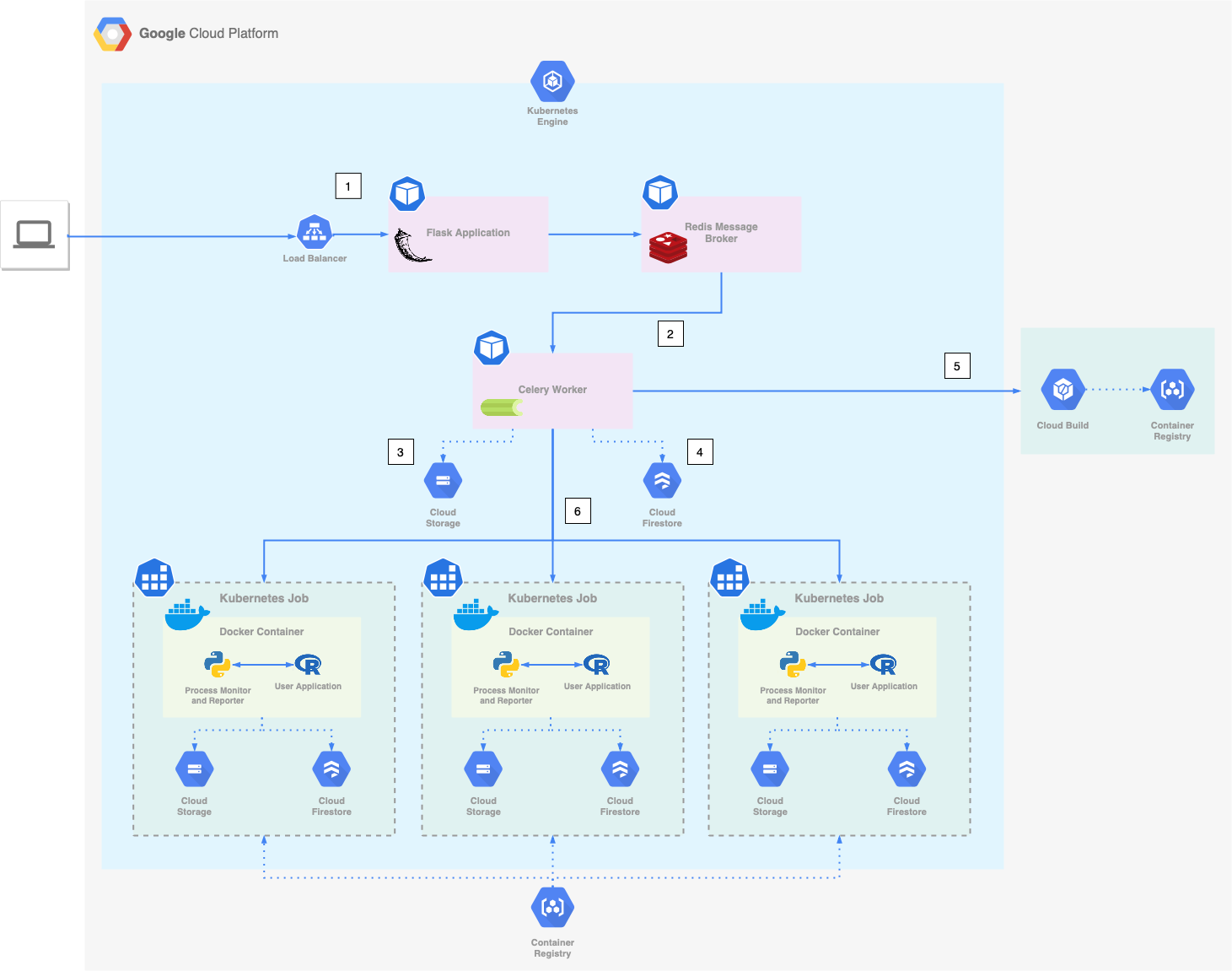}}
\caption{Technical  implementation  of  the  RE3  platform: (1) user connects to the platform, (2) the processes for file upload, reproducibility test are queued and assigned with Redis (https://redis.io) and Celery (https://docs.celeryproject.org), (3) all project files are stored on Cloud Storage, (4) metadata is stored in Cloud Firestore database, (5) Docker image building step is sent to Cloud Build service, (6) the reproducibility test is conducted as new Kubernetes job that uses pre-existing Docker containers from Container Registry. The test outputs are passed to the storage and the database.}
\label{fig:backend}
\end{figure*}

Diagram of the technical implementation of the RE3 platform is shown in Fig. \ref{fig:backend}.

\section*{Acknowledgment}

A. Trisovic is funded by the Alfred P. Sloan Foundation (grant number P-2020-13988). This research is supported by the Google Cloud credits for education. 
The authors thank Merce Crosas and Orran Krieger for enabling the collaboration between Boston University and Harvard University. Thank you to Katie Mika and Asmund Folkestad for reading the paper and their helpful comments.

\bibliographystyle{IEEEtran}
\bibliography{references}

\end{document}